\documentclass[reprint,aps,prd,twocolumn,showpacs,showkeys,superscriptaddress,floatfix,nobibnotes]{revtex4-2}

\usepackage[utf8]{inputenc}
\usepackage[T1]{fontenc}
\usepackage{graphicx}
\usepackage{bm}
\usepackage{amssymb}
\usepackage{amsmath}
\usepackage{xcolor}
\usepackage{listings}
\usepackage{url}
\usepackage{booktabs}
\usepackage{tikz}
\usetikzlibrary{shapes.geometric, arrows, positioning}
\graphicspath{{figures/}}

\newcommand{\gqgc}{\texttt{gQGC\_UFO}}

\newcommand{\skill}{\texttt{SMEFT-Pheno-Agent}}
\newcommand{\skillname}{\texttt{SMEFT-Pheno-Agent}}

\newcommand{\TeV}{\ensuremath{\,\mathrm{TeV}}}
\newcommand{\OpFgT}[1]{\ensuremath{\mathcal{O}_{gT,#1}}}  % Operator O_{gT,i}
\newcommand{\cFgT}[1]{\ensuremath{c_{gT,#1}}}             % Wilson coefficient c_{gT,i}

\begin{document}

\title{SMEFT-Pheno-Agent: a natural-language-driven AI agent for
       machine-learning-assisted Standard Model Effective Field Theory
       phenomenology}

\author{Yu-Chen Guo}
\email{ycguo@lnnu.edu.cn}
\author{Jie Wang}
\author{Ji-Chong Yang}
\email{yangjichong@lnnu.edu.cn}
\thanks{Corresponding author}
\affiliation{Center for Theoretical and Experimental High Energy Physics,\\
            Department of Physics, Liaoning Normal University, Dalian 116029, China.}
%Corresponding authors: \href{mailto:ycguo@lnnu.edu.cn}{\texttt{ycguo@lnnu.edu.cn}}, \href{mailto:yangjichong@lnnu.edu.cn}{\texttt{yangjichong@lnnu.edu.cn}}

%\date{\today}

\begin{abstract}
We present \skill{}, a Python workflow guided by a natural-language AI agent to perform machine-learning-assisted Standard Model Effective Field Theory (SMEFT) phenomenology at high-energy colliders. The software coordinates twelve automated execution phases spanning configuration intake, environment validation, event generation, machine-learning selection, statistical inference, and final audit. At each phase boundary, the agent interprets natural-language intent to generate runnable parameter files and adapter invocations required for subsequent execution. Once the detector-level events are written, the agent automatically proposes key kinematic observables alongside candidate machine-learning algorithms suited to the specific data structure and analysis objectives. All numerical calculations are delegated strictly to validated domain tools, with MadGraph5\_aMC@NLO, Pythia, Delphes generating collider simulations, and~MLAnalysis extracting features. The agent cannot modify physical parameters outside the locked configuration, and all LLM-produced artifacts, including parameter files, observable choices, algorithm selections, and prose drafts, are documented in machine-readable phase manifests prior to execution. These manifests establish complete reproducibility and audit traceability for SMEFT phenomenology studies.
\end{abstract}

%\pacs{07.05.Ml, 12.60.Fr, 13.85.-t}

\maketitle

%%%%%%%%%%%%%%%%%%%%%%%%%%%%%%%%%%%%%
\section{Introduction}
\label{sec:intro}
%%%%%%%%%%%%%%%%%%%%%%%%%%%%%%%%%%%%%

The central task of collider phenomenology is to translate a model, commonly specified through a Lagrangian and numerical assumptions, into detector-level observables and statistical statements that can be confronted with data. This translation is supported by a mature collection of specialised programs for matrix-element generation, parton-shower, detector simulation, event analysis, and inference. Executing phenomenological studies by orchestrating tools such as FeynRules~\cite{Alloul:2013bka}, FeynArts~\cite{Hahn:2000kx}, FeynCalc~\cite{Shtabovenko:2016sxi,Shtabovenko:2020gxv,Shtabovenko:2023idz}, MadGraph~\cite{Alwall:2014hca}, Pythia~\cite{Bierlich:2022pfr}, Delphes~\cite{deFavereau:2013fsa}, and MLAnalysis~\cite{Guo:2023nfu} has consequently become a standard research paradigm in HEP. Although individual tools are mature, their disparate syntaxes and usage conventions make manual workflow coordination a major bottleneck. Consequently, reliably moving from a physics request to a final result presents a fundamental orchestration challenge, where the underlying domain calculations are standardized but the handoffs between them remain brittle and manual.

This difficulty is especially pronounced in Standard Model Effective Field Theory (SMEFT) analyses. A complete SMEFT phenomenology study at a high-energy collider requires chaining together four computationally intensive steps, including matrix-element event generation, parton showering, full-coefficient scanning, and machine-learning-based event selection. Equally critical are three operationally fragile requirements: maintaining strict event-count discipline, tuning anomaly-score thresholds, and preserving audit-grade reproducibility metadata. A one-coefficient projection further requires a model-specific operator check, a Standard Model (SM) baseline, detector-level event samples, a parameter scan, an ML-defined selection, and a statistical inversion, and a small undocumented change in any of these stages can invalidate comparisons between coefficients or make a calculation difficult to reproduce. In the absence of a tool that absorbs the failure modes of every sub-task, every phase boundary forces a human-in-the-loop handoff, and the end-to-end budget of a single study easily grows from a week to a month. Existing automation frameworks reduce some of this burden, but reproducibility is not guaranteed merely by invoking the same sequence of programs; the study definition, intermediate state, and recovery decisions must also be retained.

Recent agent systems illustrate complementary ways to address these issues~\cite{Moreno:2026mqk}. ColliderAgent separates language-level reasoning from a common numerical backend for end-to-end phenomenology~\cite{Qiu:2026arch}. FERMIACC uses scaffolded, tool-assisted reasoning to formulate and test particle-theory hypotheses~\cite{FERMIACC:2026}; Dr.\,Sai organises real-world experimental analysis through specialised planning, coding, and verification components~\cite{He:2026jjb}; and RooAgent demonstrates the value of a constrained tool interface for ROOT-based analysis tasks~\cite{Desai:2026nmx}. Together with reproduction and benchmark studies~\cite{Qiu:2026vrx,Cakir:2026jev,Faroughy:2026dkj}, these works make clear that the useful role of an agent is not to replace physical modelling or numerical software, but to make the transition from a well-specified intent to validated tool calls explicit and inspectable.

\skill{} addresses this problem through a three-tier architecture based on structured task representation, deterministic computation, and evidence-chain auditing. This design clearly defines the agent's role in planning and orchestration, while delegating physical numerical calculations entirely to specialized software. The workflow supplies a twelve-phase pipeline that owns every phase boundary and takes a single bilingual plain-text intake as its only interactive stage. 
Each phase operates as a finite-state machine that produces a declared completion \emph{artifact} --- defined here as any disk-bound file or file set, such as phase manifests, data tables, trained model records, and resulting figures, generated for downstream execution or peer review. Within this framework, the AI agent is strictly confined to high-level planning and workflow orchestration at declared interfaces, taking no part in numerical derivations. Concretely, the agent ingests and validates a structured study request through a bilingual intake process, subsequently dispatching immutable configurations to specialized physics software, including MadGraph5\_aMC@NLO, Delphes, MLAnalysis, and registered machine-learning libraries. Consequently, matrix-element calculations, detector simulations, model training, and statistical inferences remain entirely under the control of these deterministic domain tools. To guarantee full traceability and reproducibility, the software maintains an append-only log that records execution contexts, fixed parameters, and all phase-completion artifacts. 

The contribution of this paper is a reproducible execution framework for one-coefficient SMEFT phenomenology. Its main features are: (i) a bilingual, validated intake that turns a research request into a locked configuration; (ii) deterministic execution phases with explicit inputs, outputs, and completion conditions; (iii) an anchor-pinned ML-selection rule that keeps the statistical comparison across coefficients consistent; and (iv) an evidence chain consisting of configuration, log, numerical outputs, and final audit. The muon-collider artifacts documented within the relevant workflow phases are a representative software demonstration, not the central physics result or a replacement for a dedicated phenomenological analysis.

The remainder of the paper is organised as follows. Section~\ref{sec:overview} describes the program architecture, its three-tier design, the configuration and log records, and the completion-artifact and audit mechanism. Section~\ref{sec:principles} sets out the computational principles that constrain the agent, the execution invariants, the anchor-pinned ML-selection rule, and the inference, limitations, and provenance record. Section~\ref{sec:phases} walks through the twelve workflow phases in order---Installation and quick start; Phases~1--6 (bilingual intake, environment validation, operator sensitivity, the SM baseline and coefficient ranges, pure event samples, and coefficient scans); Phases~7--9 (ML training and scoring, selection and fit, and the statistical interval); and Phases~10--12 (reproducibility, replay, and audit)---and includes representative artifacts from the reference execution. Section~\ref{sec:summary} concludes. Code and data availability follow in Section~\ref{sec:code}.

%%%%%%%%%%%%%%%%%%%%%%%%%%%%%%%%%%%%%
\section{Program overview}
\label{sec:overview}
%%%%%%%%%%%%%%%%%%%%%%%%%%%%%%%%%%%%%

\begin{figure*}[t]
\centering
\begin{tikzpicture}[
    intake/.style={rectangle, rounded corners, minimum width=1.8cm, minimum height=1.1cm, text centered, draw=black, fill=blue!12, font=\footnotesize\bfseries, align=center, inner sep=4pt},
    phase/.style={rectangle, rounded corners, minimum width=1.8cm, minimum height=1.1cm, text centered, draw=black, fill=green!10, font=\footnotesize\bfseries, align=center, inner sep=4pt},
    audit/.style={rectangle, rounded corners, minimum width=1.8cm, minimum height=1.1cm, text centered, draw=black, fill=red!12, font=\footnotesize\bfseries, align=center, inner sep=4pt},
    arrow/.style={thick,->,>=stealth,shorten >=2pt,shorten <=2pt},
    backarrow/.style={thick,->,>=stealth,red,shorten >=2pt,shorten <=2pt}
]
% Title
\node[font=\small\bfseries, anchor=west] at (-0.7, 3.0) {Twelve-phase execution graph};

% Row 1 (top, left-to-right): Phases 1--6
\node[intake] (p1) at (0,    1.7) {Phase 1\\Intake};
\node[phase]  (p2) at (2.4,  1.7) {Phase 2\\Env};
\node[phase]  (p3) at (4.8,  1.7) {Phase 3\\OpSens};
\node[phase]  (p4) at (7.2,  1.7) {Phase 4\\Range};
\node[phase]  (p5) at (9.6,  1.7) {Phase 5\\PureSpl};
\node[phase]  (p6) at (12.0, 1.7) {Phase 6\\Scan};

% Row 2 (bottom, right-to-left): Phase 7 rightmost, Phase 12 leftmost
\node[audit]  (p12) at (0,    0) {Phase 12\\Audit};
\node[phase]  (p11) at (2.4,  0) {Phase 11\\Replay};
\node[phase]  (p10) at (4.8,  0) {Phase 10\\MS draft};
\node[phase]  (p9)  at (7.2,  0) {Phase 9\\Bound};
\node[phase]  (p8)  at (9.6,  0) {Phase 8\\Sel+Fit};
\node[phase]  (p7)  at (12.0, 0) {Phase 7\\ML train};

% Forward arrows in row 1 (left to right)
\draw [arrow] (p1) -- (p2);
\draw [arrow] (p2) -- (p3);
\draw [arrow] (p3) -- (p4);
\draw [arrow] (p4) -- (p5);
\draw [arrow] (p5) -- (p6);

% Forward arrows in row 2 (right to left, arrowheads point left)
\draw [arrow] (p7)  -- (p8);
\draw [arrow] (p8)  -- (p9);
\draw [arrow] (p9)  -- (p10);
\draw [arrow] (p10) -- (p11);
\draw [arrow] (p11) -- (p12);

% Vertical connector: Phase 6 (south) down to Phase 7 (north), both at x=12.0
\draw [arrow] (p6.south) -- (p7.north);

% Back-signal arrow (red, from below to Phase 12)
\draw [backarrow] ([yshift=-1.1cm]p12.south)
      -- node[right=4pt, font=\scriptsize\itshape, red, xshift=2pt] {audit back-signals} (p12.south);
\end{tikzpicture}
\caption{Twelve-phase execution graph of \skill{}. Abbreviations and the
phases they label are explained in the body text immediately below.}
\label{fig:phases}
\end{figure*}
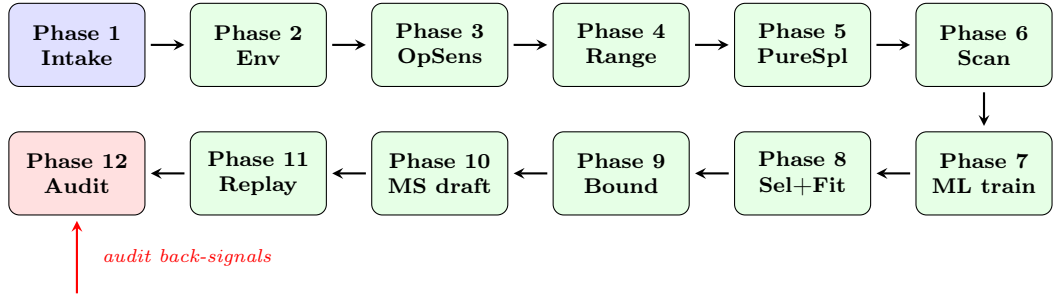

\skill{} treats a phenomenology study as a typed state transition rather than as an unstructured collection of shell commands. Figure~\ref{fig:phases} summarises the twelve phases. Each phase consumes declared inputs, writes a domain-specific completion artifact, and records its activity in \texttt{workflow.log}. A subsequent phase may begin only after the artifact from its predecessor is present and valid. This explicit handoff follows the reasoning-execution separation used in recent agent architectures~\cite{Qiu:2026arch,FERMIACC:2026} by limiting inter-stage information transfer to the minimum computational state required and providing well-defined recovery procedures for generator, detector-simulation, or remote-execution failures.

Figure~\ref{fig:phases} arranges the twelve phases as a two-row U-shape with a back-arrow from the closing audit to the next session. Phases~1--6 build the study, the data, and the numerical scan in order; Phases~7--9 select events, fit the cross-section parabola, and derive the statistical interval; \emph{Phase~10} (MS draft) assembles the manuscript evidence; \emph{Phase~11} (Replay) packages the numerical replay artifacts; \emph{Cleanup} is a sub-step of Phases~10/11; and \emph{Phase~12} (Audit) re-reads the declared artifacts for cross-phase self-consistency and emits diagnostic back-signals (red arrow) into the next session for the human reviewer. The abbreviated box labels expand as follows: \emph{Intake} = dialogue-driven study intake; \emph{Env} = environment validation; \emph{OpSens} = operator sensitivity validation; \emph{Range} = SM baseline and coefficient range; \emph{PureSpl} = pure event samples (an SM event and one pure new-physics (NP) event per claimed coefficient); \emph{Scan} = coefficient scan over the locked grid; \emph{ML train} = ML training and scoring; \emph{Sel+Fit} = selection optimisation and quadratic fit; \emph{Bound} = statistical interval derivation; \emph{MS draft} = manuscript evidence and drafting; \emph{Replay} = numerical replay package; \emph{Audit} = self-consistency audit.

Three workflow invariants follow directly from the system structure shown in Figure~\ref{fig:phases}. First, Phase~1 operates as a bilingual intake that accepts exclusively plain-text natural language, emits plain-text prompts, and avoids multiple-choice user interfaces. Second, Phases~2 to 12 execute unattended; each phase declares its completion criteria in the manifest and flags its status as \texttt{complete} only when all conditions are fulfilled, ensuring that subsequent phases proceed only after preceding artifacts are validated. Finally, cleanup functions as a sub-step within Phases~10 and 11, while Phase~12 enforces its own completion contract by validating logged artifacts against the consistency requirements and propagating diagnostic signals to subsequent sessions for review.

\subsection{Three-tier architecture and tool boundary}
\label{sec:architecture}

The first tier is the \emph{structured task representation}. It contains the process, model, beam configuration, luminosity, detector choice, execution target, event-count policy, and requested algorithms. Its purpose is to turn a human request into a finite set of values that a backend can validate and replay. The second tier is \emph{deterministic computation}. The agent dispatches only validated parameters to domain programs and does not replace their matrix-element, detector, ML, or statistical routines. The third tier is the \emph{evidence chain}, which retains the configuration, phase artifacts, command context, software revision, and audit result. This division resembles the tool-constrained approach adopted in other scientific-agent systems~\cite{Menzo:2026qrl,Lucente:2026kgh}, while retaining a deliberately small scope: one configured SMEFT workflow rather than open-ended theory generation or experimental reconstruction.

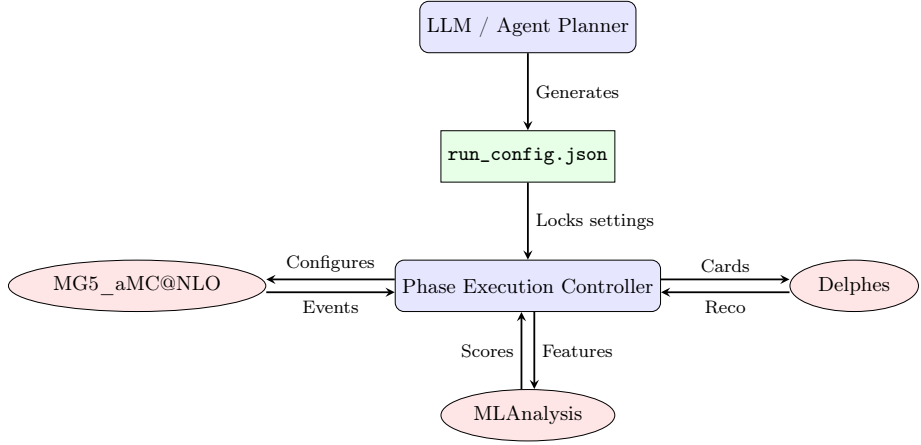
\begin{figure*}[t]
\centering
\scalebox{0.85}{
\begin{tikzpicture}[node distance=1.2cm,
    process/.style={rectangle, rounded corners, minimum width=2.5cm, minimum height=0.8cm, text centered, draw=black, fill=blue!10, font=\small},
    artifact/.style={rectangle, minimum width=2cm, minimum height=0.8cm, text centered, draw=black, fill=green!10, font=\small},
    tool/.style={ellipse, minimum width=2cm, minimum height=0.8cm, text centered, draw=black, fill=red!10, font=\small},
    arrow/.style={thick,->,>=stealth}]

    \node (llm) [process] {LLM / Agent Planner};
    \node (config) [artifact, below=of llm] {\texttt{run\_config.json}};
    \node (controller) [process, below=of config] {Phase Execution Controller};
    \node (mg5) [tool, left=of controller, xshift=-0.8cm] {MG5\_aMC@NLO};
    \node (delphes) [tool, right=of controller, xshift=0.8cm] {Delphes};
    \node (ml) [tool, below=of controller] {MLAnalysis};

    \draw [arrow] (llm) -- node[anchor=west, font=\footnotesize] {Generates} (config);
    \draw [arrow] (config) -- node[anchor=west, font=\footnotesize] {Locks settings} (controller);
    \draw [arrow, transform canvas={yshift=1mm}] (controller) -- node[anchor=south, font=\footnotesize] {Configures} (mg5);
    \draw [arrow, transform canvas={yshift=-1mm}] (mg5) -- node[anchor=north, font=\footnotesize] {Events} (controller);

    \draw [arrow, transform canvas={yshift=1mm}] (controller) -- node[anchor=south, font=\footnotesize] {Cards} (delphes);
    \draw [arrow, transform canvas={yshift=-1mm}] (delphes) -- node[anchor=north, font=\footnotesize] {Reco} (controller);

    \draw [arrow, transform canvas={xshift=1mm}] (controller) -- node[anchor=west, font=\footnotesize] {Features} (ml);
    \draw [arrow, transform canvas={xshift=-1mm}] (ml) -- node[anchor=east, font=\footnotesize] {Scores} (controller);
\end{tikzpicture}
}
\caption{Data flow diagram of \skillname{}. The LLM generates the immutable configuration file, which acts as the central contract. The execution controller then orchestrates the deterministic numerical tools, passing data between them without further language-model intervention.}
\label{fig:dfd}
\end{figure*}

Figure~\ref{fig:dfd} illustrates the data flow across these tiers. Operating at the orchestration layer, the LLM interprets natural-language intent at each phase boundary to generate the runnable parameter files and adapter invocations required for subsequent execution. Furthermore, once physics samples become available, it recommends relevant kinematic observables alongside candidate machine-learning algorithms tailored to the data structure and analysis objectives. Tool adapters then compose MG5 configurations, prepare Delphes cards, convert generated events, execute machine-learning routines, and extract declared outputs, ensuring that validated physics software remains the sole producer of numerical results. To maintain integrity, the agent cannot alter physical parameters via ad-hoc command lines after locking the configuration. Moreover, all LLM-generated artifacts, including parameter files, observable selections, algorithm proposals, and prose drafts, are recorded in phase manifests prior to execution, enabling deterministic replay of the entire calculation without re-invoking the LLM.

Crucially, because every Monte Carlo simulation and detector response is executed locally by the same specialist software as in a manual run, the raw numerical execution time of each individual phase is identical to a manual execution. The performance gain is therefore not in raw compute time, but in the orchestration layer that surrounds the computation. The workflow automates four classes of activity that are otherwise manual, error-prone, and costly in researcher time: (i)~the handoffs between MG5, MLAnalysis, and the analysis scripts with ML, which absorb the per-tool interface burden and the recovery semantics at every phase boundary; (ii)~the data-driven selection of the kinematic feature representation and the candidate ML algorithm family appropriate to the detector-level event topology; (iii)~the per-algorithm threshold scan and automated comparison of the resulting $S_{\mathrm{stat}}$ intervals, with the per-operator comparison retained in the phase manifest; and (iv)~the assembly of a manuscript draft from the validated manifests, numerical tables, and plots, with every LLM-authored sentence grounded in a specific manifest entry. Together these eliminate the cumulative cost of human intervention, the cognitive burden of learning and troubleshooting each tool's interface, and the risk of silent configuration drift between phases.

\subsection{Configuration and log records}
\label{sec:records}

The configuration file \texttt{run\_config.json} is the single source of truth for a run. It serialises physics inputs together with operational inputs that are often omitted from a manual analysis note: executable locations, execution target, detector card, polarisation, requested algorithms, and the event-count policy. In particular, the event counts used by the sensitivity, baseline, and scan phases are written once during intake and are not silently resized later. The file records values rather than conversational interpretations, allowing the numerical execution to be replayed without depending on the original prompt.

The append-only \texttt{workflow.log} complements the configuration. Each JSON-line record identifies the phase, event, command context, working directory, timestamp, software version, and revision identifier. Recovery attempts are logged as such, rather than being overwritten by a successful later execution. These records provide a compact persistent state, avoiding a dependence on long conversational histories while preserving the provenance needed by a later phase or a human reviewer. Together, the two files distinguish a planned calculation from the calculation that was actually carried out.

\subsection{Completion artifacts and audit}
\label{sec:contract}

Workflow completion is defined by domain-specific artifacts, spanning environment manifests, event samples, trained-model records, and manuscript evidence sets. Shared parameters across these artifacts, such as the global operator key, algorithm selection, threshold, and scan label, act as explicit consistency checks. Phase~12 performs a self-consistency audit by confirming that all declared artifacts exist, parse properly, and match across these shared identifiers. Unresolved numerical roots, failed recovery attempts, and missing data are represented explicitly rather than filled with fallback values. While such an audit provides evidence of execution integrity, it does not by itself establish the physical validity of a given model, detector card, or statistical method.

%%%%%%%%%%%%%%%%%%%%%%%%%%%%%%%%%%%%%
\section{Computational principles}
\label{sec:principles}
%%%%%%%%%%%%%%%%%%%%%%%%%%%%%%%%%%%%%

\subsection{Interaction boundary and constrained agent role}

The agent has two responsibilities: it collects a complete study specification and orchestrates the phase transitions prescribed by that specification. It does not derive matrix elements, replace detector simulation, or infer unrecorded physical assumptions. Numerical results are produced by the registered specialist tools and are accepted only when the phase artifact satisfies its validation rules. This division keeps language-level flexibility at the interface while retaining deterministic, independently inspectable computation in the numerical core.

Phase~1 functions as the only \emph{intake-style conversation}.
Each bilingual prompt is checked before requesting next fields, and once a valid \texttt{run\_config.json} file is generated, the agent transitions from a prompt-response dialogue to an artifact-driven operational mode.
From Phase~2 onward, although the LLM continues to participate at every phase boundary, the interactive dialogue is replaced by three deterministic actions. First, the agent reads the manifest from the preceding phase. Second, it makes data-aware decisions strictly within the constraints defined in the locked configuration, such as selecting feature schemas, proposing candidate machine-learning algorithms, defining scan grids, and drafting manuscript prose grounded in manifest data. Third, it emits the input for the next phase as machine-readable JSON or a runnable script. The agent cannot modify physical parameters outside the locked configuration, and every LLM-generated decision is recorded in the phase manifest before executing the corresponding numerical step. This design, which pairs flexible natural language at the interface with validated deterministic code in the numerical core and a structured artifact chain between them, aligns with constraint patterns in modern scientific agent systems~\cite{He:2026jjb,Desai:2026nmx}. Expressive language captures human intent effectively, whereas repeatable scientific execution demands a constrained tool interface and explicit validation boundaries.

\subsection{Execution invariants and recovery semantics}

Several invariants protect comparisons across a long calculation. The locked event counts prevent a costly scan from being silently reduced after a partial failure. Phase~3 completes the full model-parameter loop before deciding which operators are sensitive, so a single zero cross section cannot prematurely end the study. The global operator identifier, rather than a UFO-internal position, is propagated through event files, ML scores, fits, and constraint records. Recovery operations may repair a missing conversion or rerun a failed backend job, but they must append a new log record and preserve the original status. Thus, recovery changes the execution state in a visible way instead of retrospectively changing the evidence.

\subsection{Feature schema and anchor-pinned ML selection}
\label{sec:anchor}

The machine-learning (ML) input itself serves as a declared artifact. A feature schema specifies particle selection, ordering rules, feature dimensions, preprocessing pipelines, random seeds, training splits, and file paths for all fitted models. This record is essential because an ML score lacks portable meaning without its explicit reconstruction and preprocessing context. In practice, after inspecting detector-level final-state content and event topology, the LLM proposes the schema by selecting particle objects, ordering conventions, preprocessing routines, and a constrained set of candidate ML algorithms suited to the data structure and analysis goals. Bound by the configuration and pre-registered transformation routines, this proposal is logged in the schema manifest prior to training. Treating the feature schema as a first-class Phase~7 artifact rather than an ad-hoc decision guarantees that a reviewer can reconstruct the exact preprocessing steps and feature definitions behind any reported score.

Machine-learning performance depends on both the chosen coefficient and the score threshold. Independent selection of algorithms and thresholds for each coefficient improves individual intervals at the cost of cross-operator comparability. Conversely, enforcing a universal operating point across all operators yields sub-optimal limits for coefficients with distinct kinematic profiles. To make this trade-off explicit, the workflow supports two \emph{selection scopes} declared in the configuration:
\begin{itemize}
\item \texttt{anchor\_operator\_only} (the default used in the reference run): the workflow evaluates a fixed threshold grid on the anchor and selects the algorithm--threshold pair that minimises the chosen anchor criterion. The selected pair is written to the evidence record and propagated unchanged to the remaining one-coefficient scans, so the headline intervals are taken with a single locked operating point.
\item \texttt{per\_operator\_best}: the workflow fits every (operator, algorithm) pair and reports, for each operator, the algorithm that yields the narrowest $S_{\mathrm{stat}}$ interval on that operator's own scan.
\end{itemize}
In both scopes the per-operator comparison data is preserved in the phase manifest (\texttt{comparison\_only\_intervals}), so an external user can switch scopes after the fact by reading the manifest and applying a different aggregation. The default is a workflow policy, not a claim that one classifier is universally optimal or that a classifier trained at one reference point yields an optimal selection for all possible EFT scenarios.

\subsection{Inference, limitations, and provenance}
\label{sec:inference}

Following selection, the post-cut cross section is parameterised by a quadratic fit,
\begin{equation}
    \sigma^{\mathrm{cut}}(f)=\sigma^{\mathrm{cut}}_{\mathrm{SM}}+
f\sigma_{\mathrm{int}}+f^2\sigma_{\mathrm{NP}},\label{eq:csfit}
\end{equation}
where $\sigma^{\mathrm{cut}}_{\mathrm{SM}}$ is fixed to the Standard Model prediction. We express the dimension-eight Wilson coefficient $f=\cFgT{i}$ in $\mathrm{TeV}^{-4}$, corresponding to the conventional $1/\Lambda^4$ suppression. Dimensional consistency then requires $\sigma_{\mathrm{int}}$ and $\sigma_{\mathrm{NP}}$ to be quoted in $\mathrm{pb}\,\mathrm{TeV}^{4}$ and $\mathrm{pb}\,\mathrm{TeV}^{8}$, respectively, while $\sigma^{\mathrm{cut}}$ is given in pb. Values stored internally in the UFO parameter card in $\mathrm{GeV}^{-4}$ are converted using $1~\mathrm{GeV}^{-4}=10^{12}~\mathrm{TeV}^{-4}$ before presentation.

To determine constraints for specified significance levels, the workflow inverts the asymptotic Poisson significance~\cite{Cowan:2010js},
\begin{equation}
    S_{\mathrm{stat}}=\sqrt{2\left[(N_b+N_s)\ln\left(1+\frac{N_s}{N_b}\right)-N_s\right]},
    \label{eq:asimov}
\end{equation}
where $N_b$ and $N_s$ are the post-selection background and signal yields for a given luminosity.

Failed scan points, non-finite fits, or missing real roots are flagged as technical limitations rather than converted into physical bounds. Because the resulting intervals depend strictly on the single-coefficient hypothesis, detector simulation, feature selection, and statistical approximations, the output provenance records both the derived limits and any execution failures. These bounds should be interpreted as conditional algorithm outputs rather than global SMEFT fits or experimental likelihoods.

%%%%%%%%%%%%%%%%%%%%%%%%%%%%%%%%%%%%%
\section{Workflow phases}
\label{sec:phases}
%%%%%%%%%%%%%%%%%%%%%%%%%%%%%%%%%%%%%

\subsection{Installation and quick start}
\label{sec:quickstart}

The workflow is distributed as a portable agent skill following the Agent Skills open standard. The reference implementation uses a \texttt{SKILL.md} file and can be deployed on compatible agent runtimes that support this standard or an equivalent local skill mechanism. Representative examples include Codex, Claude Code, Kimi Code, MiniMax Code, and QoderWork. This list is illustrative rather than exhaustive.

The skill is installed by placing the repository at the runtime's skill discovery path, and becomes active in the next session without restarting the runtime:
\begin{lstlisting}[language=bash]
# Clone from Gitee (primary) or GitHub mirror (see Section \ref{sec:code})
git clone https://github.com/NBAlexis/SMEFT-Pheno-Agent.git
# Codex
ln -s "$(pwd)/SMEFT-Pheno-Agent" ~/.codex/skills/SMEFT-Pheno-Agent-main
# Claude Code: ~/.claude/skills/   |   Codex/OpenCode/Kimi Code: analogous paths
# On Windows: replace ~/.%AGENTNAME%/skills/ with %USERPROFILE%\.%AGENTNAME%\skills\
\end{lstlisting}

The MG5 and MLAnalysis toolchains are not required to be pre-installed. After the skill is loaded, the user starts a new session under any of the supported runtimes and requests the study in natural language, e.g.\ an SMEFT phenomenology study on $pp\to t\bar t$ at 13\TeV.
The runtime auto-loads the skill via description match, and Phase~1 begins. The reference execution documented in this paper uses a muon-collider process $\mu^+\mu^-\to\nu_\mu\bar\nu_\mu jj$ at $\sqrt{s}=10$\TeV (Table~\ref{tab:reference-setup}) for demonstration. Other processes such as $pp\to t\bar t$ are equally supported through the same interface.
Phase~1 is the only interactive stage. It sequentially prompts for nine input parameters---including software paths, target physics processes, UFO models, beam energies, luminosities, and detector card selection---before compiling them into \texttt{run\_config.json}. The execution target can be local, \texttt{ssh user@host:/path/to/MG5}, or \texttt{wsl2:/path/to/MG5}; in SSH/WSL2 mode the path refers to the target Linux filesystem, not the Windows host. If MG5 or MLAnalysis are not present on the target, the user can opt to let the skill install them via \texttt{scripts/install\_mg5.py} and \texttt{scripts/install\_cutexperiment.py}.

Once \texttt{run\_config.json} is written, the event counts are locked and Phases 2--12 run autonomously: SM/NP event generation through MG5, Delphes/root2lhco conversion, MLAnalysis feature extraction, ML event selection, Wilson-coefficient scanning, expected-limit computation, manuscript generation, MG5 process folder cleanup, replay packaging, and self-consistency auditing. The skill does not ask the user whether to continue, does not downsize the numerical budget for speed, and does not replace a missing LaTeX build with an HTML fallback. Long MG5/Delphes runtimes (hours to weeks) are expected and are handled with durable execution and status files; on Windows hosts, the long-running phases must be dispatched through WSL2 or SSH rather than native Windows local execution.

This separation prevents late conversational changes from modifying a calculation after its numerical budget has been fixed.

\subsection{Phase 1: bilingual study intake}
\label{sec:phase1}

Phase~1 is the only conversational phase. The agent gathers the execution target, MG5 and MLAnalysis locations, physics process, UFO model list, collider energy or beam energies, luminosity, detector card, and optional ML and polarisation settings. The dialogue follows a fixed sequence of bilingual prompts. Each response is validated before the next question is asked; invalid paths, incompatible final states, or malformed numeric inputs are re-asked with the relevant constraint. After the final confirmation, the agent writes \texttt{run\_config.json} and locks the event counts. This is the last point at which the study definition can be changed through dialogue.

For the reference execution used to illustrate the phase artifacts below, the locked configuration is summarised in Table~\ref{tab:reference-setup}. The table is an example of a Phase~1 deliverable: it records the study definition rather than a post-hoc interpretation of the calculation. Requested event budgets are immutable; realised LHCO event counts and recovery attempts are recorded later in the relevant phase manifests.

\begin{table*}[htb]
\centering
\caption{Phase~1 configuration excerpt for the reference run.}
\label{tab:reference-setup}
\begin{tabular}{ll}
\toprule
\multicolumn{1}{c}{Quantity} & \multicolumn{1}{c}{Locked value} \\
\midrule
Process & $\mu^+\mu^-\to\nu_\mu\bar\nu_\mu jj$ \\
Centre-of-mass energy & $\sqrt{s}=10$~TeV \\
Integrated luminosity & $10\ \mathrm{ab}^{-1}$ \\
Configured polarisation tuple & $(-1,+1)$ \\
Model & \gqgc{} \\
Detector simulation & \texttt{delphes\_card\_MuonColliderDet.tcl} \\
Final-state requirement & $N_{\mathrm{jet}}\geq 2$ \\
Locked event budgets (events) & $N_{\rm SM}=1\times10^6$, $N_{\rm NP}=10^5$, $N_{\rm scan}=10^6$ \\
\bottomrule
\end{tabular}
\end{table*}

\subsection{Phase 2: environment validation}

Phase~2 probes the configured execution target and validates the paths and versions of MG5, Delphes, MLAnalysis, Python dependencies, detector card, and write locations. It writes an environment manifest that identifies the available tools and their probe outcomes. If a prerequisite is unavailable, the phase reports the concrete failure instead of allowing later phases to produce ambiguous partial output.

\subsection{Phase 3: operator sensitivity validation}

For each configurable coefficient in each requested UFO model, Phase~3 performs a small, locked-event-count generation and reads the generator cross section. The phase completes the full coefficient loop before deciding whether any coefficient is usable. Non-zero coefficients are recorded with stable global operator identifiers, and the first suitable entry becomes the default anchor. The resulting \texttt{operator\_claims.json} makes the scope of the later scan explicit. In the reference run, all eight \gqgc{} coefficients \OpFgT{0}--\OpFgT{7} (coded as \texttt{FgT0}--\texttt{FgT7}) are claimed, and the first global key, \OpFgT{0}, is recorded as the anchor. This identity is carried unchanged into the event, score, fit, and constraint artifacts.

\subsection{Phase 4: SM baseline and coefficient ranges}

Phase~4 generates an SM baseline and estimates a scan magnitude separately for each claimed coefficient. The purpose is operational rather than interpretive: the range should give a numerically usable coefficient grid for the subsequent scan. The phase writes the baseline result and \texttt{coefficient\_ranges.json}, including the proposed magnitude and its provenance. For the reference \OpFgT{0} artifact, the proposed magnitude is $2.1\times10^{-2}~\mathrm{TeV}^{-4}$, defining the eleven-point interval $[-2.1,2.1]\times10^{-2}~\mathrm{TeV}^{-4}$ used by Phase~6. This range is an operational scan choice, not a coefficient limit.

\subsection{Phase 5: pure event samples}

The workflow creates one SM event and one pure-NP event per claimed coefficient, adhering strictly to the locked event counts. These generator outputs are processed through detector simulation and converted into the target event representation for MLAnalysis. By preserving loaded event counts and recovery logs alongside the samples, downstream ML pipelines avoid assuming full event availability without verification. For the reference run, the \OpFgT{0} anchor event is generated at the model-default value $\cFgT{0}=1~\mathrm{TeV}^{-4}$. This event serves as a training benchmark to establish SM--NP discrimination rather than acting as an explicit constraint.

\subsection{Phase 6: coefficient scans}

For every claimed coefficient, Phase~6 evaluates the configured grid while holding the remaining coefficients fixed. Each point has a stable label and a separate execution record. The phase produces a scan manifest containing the coefficient values, output locations, event-count metadata, and recovery status. When a point cannot be regenerated, its state is explicitly recorded in the manifest, allowing a subsequent fit to determine whether to exclude the point or leave the entire coefficient unconstrained.

\subsection{Phase 7: ML training and scoring}

Phase~7 establishes the feature schema and trains a candidate set of machine-learning algorithms. Upon inspecting the detector-level final-state topology, observable set, and event size, the LLM aligns the feature representation and algorithm selection directly with the user's overarching analysis goals.
The proposal is constrained by a feature schema declared in the configuration and by a registered admissible set of algorithms; the LLM is free to choose within these constraints but cannot introduce algorithms or observables that lie outside the declared envelope. Each candidate is then trained, and the fitted models, scalers when required, score files, and a comparison manifest are written. No algorithm is discarded at this stage on the basis of an interim metric; this preserves the comparison needed by the anchor-pinned selection rule in Phase~8. The \OpFgT{0} training scores therefore constitute a Phase~7 artifact, whereas choosing an operating point from those scores is deferred to Phase~8.

\subsection{Phase 8: selection optimisation and fits}

Phase~8 scans the configured score thresholds for every trained algorithm. For the anchor, it evaluates the selection criterion and chooses one algorithm and threshold pair. The phase then applies that pair consistently to the coefficient scans and fits Eq.~\eqref{eq:csfit} where the available points permit. Threshold scans, fitted coefficients, fit status, and unavailable roots are written as machine-readable evidence rather than being retained only in figures.

\begin{figure*}[htbp]
\centering
\includegraphics[width=0.48\linewidth]{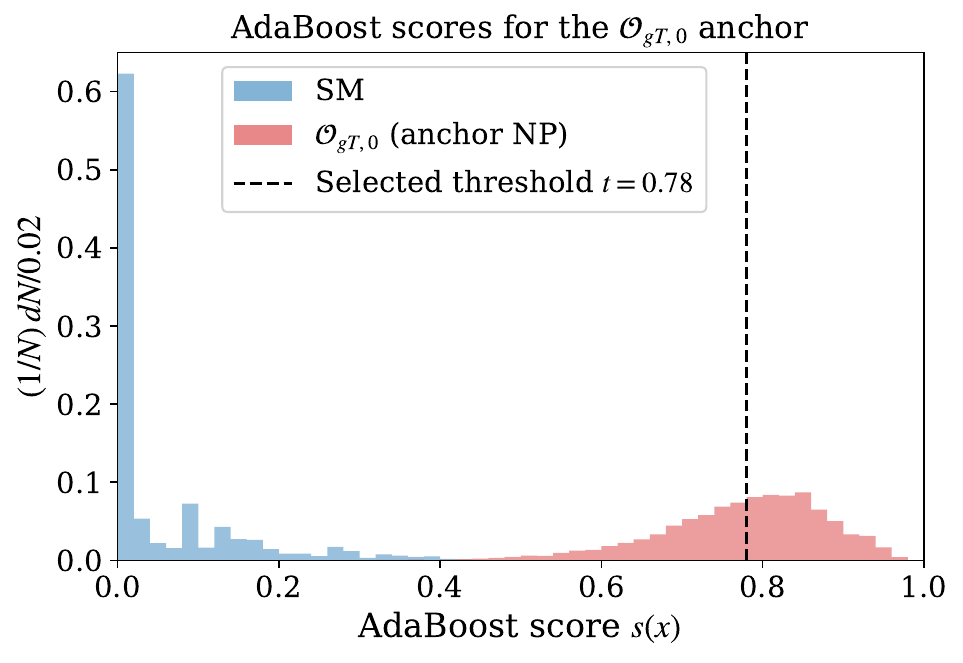}\hfill\includegraphics[width=0.48\linewidth]{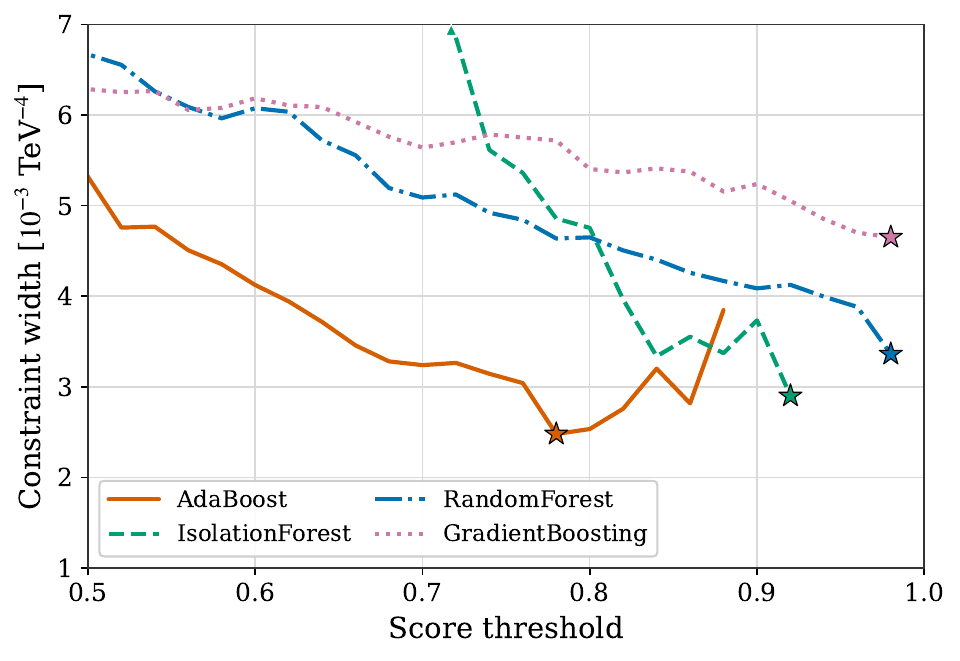}
\caption{Phase~7--8 ML evidence for the \OpFgT{0} anchor. Left: AdaBoost scores for the SM and the pure-NP \OpFgT{0} training reference ($\cFgT{0}=1~\mathrm{TeV}^{-4}$); the dashed line is the Phase~8 selected threshold. Right: Width of the finite anchor $S_{\rm stat}=2$ interval as a function of threshold for the four selectors with valid, consistently oriented stored scores. Stars mark the selected operating points, and curves terminate when the corresponding $f=0$ yield vanishes. }
\label{fig:phase8-selection}
\end{figure*}

Figure~\ref{fig:phase8-selection} shows the linked Phase~7--8 evidence for the reference anchor. The left panel displays the calibrated AdaBoost scores for the SM and the pure-NP \OpFgT{0} training sample. Each algorithm uses a fixed min--max map fitted on the training reference and reused unchanged for every downstream event, so thresholds have a consistent $[0,1]$ meaning. The right panel selects AdaBoost at $t=0.78$ because this pair gives the narrowest finite anchor $S_{\rm stat}=2$ interval. The selected pair is then locked for every coefficient scan rather than retuned coefficient by coefficient.

Table~\ref{tab:ml-comparison} summarizes the Phase~8 comparison of registered ML algorithms on the \OpFgT{0} anchor at $S_{\rm stat}=2$. AdaBoost is selected because it gives the narrowest finite interval. Under the default \texttt{anchor\_operator\_only} scope, this algorithm and operating point are then locked for the downstream fit and constraint calculation.

\begin{table*}[htb]
\centering
\caption{Phase~8 ML algorithm comparison for the \OpFgT{0} anchor at $S_{\rm stat}=2$. Constraint strength is measured by the interval width. Under the default \texttt{anchor\_operator\_only} scope, AdaBoost at $t=0.78$ is selected for the downstream analysis.}
\label{tab:ml-comparison}
\begin{tabular}{lccc}
\toprule
Algorithm & Optimal threshold & Constraint width  & Selected \\
 & & ($10^{-3}~\mathrm{TeV}^{-4}$) & \\
\midrule
AdaBoost & 0.78 & 2.479 & \checkmark \\
Isolation Forest & 0.92 & 2.899 &  \\
Random Forest & 0.98 & 3.361 &  \\
Gradient Boosting & 0.98 & 4.651 &  \\
\bottomrule
\end{tabular}
\end{table*}

\begin{figure}[t]
\centering
\includegraphics[width=1.0\linewidth]{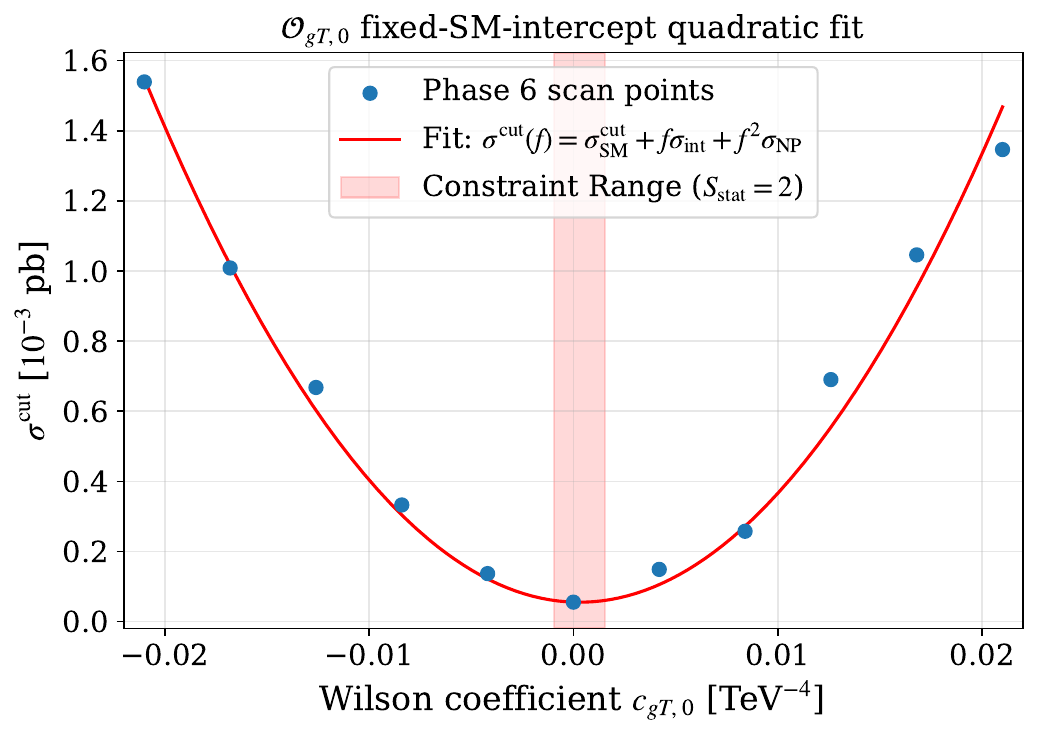}
\caption{Phase~8 fixed-SM-intercept quadratic fit for the selected \OpFgT{0} scan. The shaded band is the conditional Phase~9 $S_{\rm stat}=2$ interval.}
\label{fig:phase8-fit}
\end{figure}

The candidate operating points are evaluated with a common $f=0$ event, which corresponds to the SM with the vanishing NP contribution. If the provisional optimum leaves no background event after the cut, the workflow supplements this event with an independent coefficient-zero event and repeats the complete threshold sweep before finalizing the selection. The reference result shown here is obtained after this recovery step. Producing the narrowest supported anchor interval while retaining events in the selected $f=0$ event, AdaBoost at $t=0.78$ is locked for subsequent fits and constraint calculations.The selected pair is then locked for every coefficient scan rather than retuned coefficient by coefficient. Figure~\ref{fig:phase8-fit} shows the eleven \OpFgT{0} scan points and the corresponding fixed-intercept quadratic fit, while Table~\ref{tab:post-cut-sm-intercept} records the coefficients used in that fit.

\begin{table}[h]
\centering
\caption{Coefficients of the fixed-intercept quadratic fit in Eq.~\eqref{eq:csfit} for the illustrative \OpFgT{0} analysis. The three numerical columns are quoted in $\mathrm{pb}$, $\mathrm{pb}\,\mathrm{TeV}^{4}$, and $\mathrm{pb}\,\mathrm{TeV}^{8}$, respectively.}
\label{tab:post-cut-sm-intercept}
\begin{tabular}{lccc}
\toprule
Operator & $\sigma^{\rm cut}_{\rm SM}$ & $\sigma_{\rm int}$ & $\sigma_{\rm NP}$ \\
\midrule
\OpFgT{0} & $5.55\times10^{-5}$ & $-1.92\times10^{-3}$ & $3.29$ \\
\bottomrule
\end{tabular}
\end{table}

\subsection{Phase 9: statistical intervals}
\label{sec:phase9}

Using the Phase~8 selection record, Phase~9 evaluates requested significance levels through Eq.~\eqref{eq:asimov} and generates a coefficient-interval table. This table documents the selected algorithm, operating threshold, fit inputs, and explicit reasons for any unavailable intervals, thereby directly linking each numerical result to its underlying selection decision. Table~\ref{tab:phase9-fgt0-intervals} presents the \OpFgT{0} endpoint for the reference chain. These intervals remain strictly conditional on the single-coefficient hypothesis, the configured detector simulation, and the chosen AdaBoost model at $t=0.78$.

\begin{table}[h]
\centering
\caption{Illustrative \OpFgT{0} sensitivity ranges from the selected AdaBoost model. The same Phase~8 selection is used for every scan point.}
\label{tab:phase9-fgt0-intervals}
\begin{tabular}{cc}
\toprule
$S_{\rm stat}$ & \cFgT{0} ($10^{-3}~\mathrm{TeV}^{-4}$) \\
\midrule
2 & $[-0.948,\ 1.531]$ \\
3 & $[-1.218,\ 1.800]$ \\
5 & $[-1.655,\ 2.237]$ \\
\bottomrule
\end{tabular}
\end{table}

The reference execution is intended to demonstrate integration and provenance, not an experimentally complete SMEFT projection. It does not include a global fit, correlated coefficients, detector systematics, theory uncertainties, or an experimental likelihood. These limitations are retained as part of the interpretation record rather than being hidden by the automated interval output.

\subsection{Phases 10--12: reproducibility, replay, and audit}
\label{sec:phase12}

Phase~10 assembles the validated configuration, phase manifests, numerical tables, plots, and audit metadata into the manuscript evidence set. The LLM may draft prose and arrange manuscript elements, but every numerical statement must be traceable to a validated manifest entry. Missing results and audit warnings remain explicit, and the software and document-build environments are recorded so that the manuscript can be regenerated from the evidence set. The complete evidence hierarchy and file-level provenance are provided with the public reproducibility package rather than reproduced in the article.

Phase~11 produces a self-contained replay package containing the executable workflow, frozen configuration, ordered commands, random seeds, checksums, data documentation, and validation results. It regenerates the numerical results from Phases~2--9 in a fresh output directory and does not use the original derived outputs as production inputs. The detailed directory layout, schemas, and entry-point names are documented in the repository.

Phase~12 performs a deterministic self-consistency audit of the complete artifact chain. It checks artifact existence and parseability, cross-phase operator and selection identifiers, event-count discipline, and provenance completeness. This audit confirms that the workflow executed the declared study definition rather than establishing the physical validity of the underlying model, detector assumptions, feature choices, or statistical approximations. While an LLM may explain a failed check, it cannot influence the binary pass or fail decision, and all diagnostic outputs are recorded prior to any corrective action.

For the reference execution, all declared artifacts are present, the stored files parse, operator identifiers and the selected AdaBoost operating point agree across phases, and the locked event counts match the manifests. Interpretive limitations, including the one-coefficient hypothesis and the absence of systematic uncertainties, are reported separately from workflow-integrity failures. Detailed audit results and machine-readable failure diagnostics are included in the reproducibility package.

%%%%%%%%%%%%%%%%%%%%%%%%%%%%%%%%%%%%%
\section{Summary}
\label{sec:summary}
%%%%%%%%%%%%%%%%%%%%%%%%%%%%%%%%%%%%%

In this paper, we have presented \skill{}, an AI-agent workflow designed for reproducible, ML-assisted SMEFT phenomenology.
Its architecture integrates user intent with execution by decoupling flexible natural-language orchestration from deterministic domain computation and evidence-chain auditing.
While the agent interprets intent, regulates phase transitions, and logs calculation records, specialized domain software remains exclusively responsible for event generation, detector simulation, machine-learning estimation, and statistical inference.

The workflow incorporates a bilingual intake interface, a strict event-count policy, immutable phase artifacts, an anchor-pinned ML selection strategy, and an end-to-end self-consistency audit.
These design choices directly address a core challenge identified in emerging scientific-agent benchmarks: useful automation must be accompanied by transparent assumptions, inspectable tool calls, and a clear distinction between generated evidence and unvalidated interpretation~\cite{Qiu:2026arch,Menzo:2026qrl,Qiu:2026vrx,Faroughy:2026dkj}.

Demonstrated here via a muon collider benchmark, \skill{} provides a controlled operational framework where AI handles process orchestration while researchers retain full control over physical setups, validation, and domain interpretation. Crucially, because the agent relies on MadGraph as its numerical engine, this architecture naturally scales to any supported collider environment. While the current implementation focuses on single-operator SMEFT scenarios, future developments will extend the deterministic backend to multi-operator configurations for global fits, as well as incorporate comprehensive systematic uncertainty treatments.

\section{Code and Data Availability}
\label{sec:code}
%%%%%%%%%%%%%%%%%%%%%%%%%%%%%%%%%%%%%

The complete source code of SMEFT-Pheno-Agent is freely available under the MIT license at the project repository. The source code of \skill{}, installation scripts, and documentation are available at \url{https://github.com/NBAlexis/SMEFT-Pheno-Agent.git}. A gitee mirror is maintained at \url{https://gitee.com/NBAlexis/automated-smeft-ml-pheno}.
The repository includes installation scripts, the full workflow implementation, configuration examples, and documentation.
Reference execution artifacts (configuration files, event samples, ML models, and constraint tables) from the muon-collider example presented in Section \ref{sec:phases} are available in the repository's examples/ directory.
The workflow requires MadGraph5\_aMC@NLO and Delphes to be pre-installed or allows automatic installation via provided scripts. Long-running phases (event generation) may require hours to weeks depending on the configured event budget. Windows users must use WSL2 or SSH execution mode for Monte Carlo simulations.

\section*{Statements and Declarations}

\subsection*{Funding}
This work was supported in part by the National Natural Science Foundation of China under Grants Nos.~12575106 and 12147214 and by the Basic Research Projects of Universities in Liaoning Province under Grant No.~LJ212510165024.

\subsection*{Competing interests}
The authors declare that they have no financial or non-financial interests that are directly or indirectly related to this work.

\subsection*{Author contributions}
\textbf{Yu-Chen Guo}: Conceptualization, Investigation, Validation, Writing -- original draft, Writing -- review and editing, Funding acquisition. \textbf{Jie Wang}: Validation, Visualization. \textbf{Ji-Chong Yang}: Conceptualization, Software, Investigation, Methodology, Writing -- review and editing, Funding acquisition.

\bibliographystyle{apsrev4-2}
\bibliography{references}

\end{document}